\documentclass[9pt,twocolumn,twoside]{pnas-new}
\templatetype{pnasinvited} 

\title{Are GCMs obsolete?}

\author[1,2]{V. Balaji}
\author[3]{Fleur Couvreux} 
\author[4]{Julie Deshayes}
\author[5]{Jacques Gautrais}
\author[6]{Fr\'ed\'eric Hourdin}
\author[7]{Catherine Rio}

\affil[1]{Cooperative Institute for Modeling the Earth System, Princeton University, New Jersey, USA}
\affil[2]{Laboratoire de Sciences de Climat et Environnement, Saclay, France}
\affil[3]{CNRM, University of Toulouse, Meteo-France, CNRS, Toulouse, France}
\affil[4]{Sorbonne Universit\'es (UPMC, Univ Paris 06)-CNRS-IRD-MNHN, LOCEAN Laboratory, Paris, France}
\affil[5]{Centre de Recherches sur la Cognition Animale (CRCA), Centre de Biologie Intégrative (CBI), Université de Toulouse; CNRS, UPS, France}
\affil[6]{LMD-IPSL, Sorbonne University, CNRS, Paris, France}
\affil[7]{CNRM, University of Toulouse, Meteo-France, CNRS, Toulouse, France}

\leadauthor{Balaji} 

\significancestatement{General circulation models (GCMs) are numerical representations of the Earth system and its constituents, used to understand how the climate evolves in response to external perturbations, such as natural or human-caused changes to the atmosphere, ocean and biosphere. The models used to study how the climate responds to perturbations do not resolve all phenomena of interest. They are also too expensive to study all the possible climate pathways in the past, present, or future. But new methods built around GCMs can learn from models that can resolve some of these features, while also making sure that the models used to study the pathways are physically sound.}

\authorcontributions{VB led the writing team. All authors share equal responsibility for the conception and content of the article.}
\authordeclaration{No competing interests.}
\equalauthors{}
\correspondingauthor{\textsuperscript{2}To whom correspondence should be addressed. E-mail: balaji@princeton.edu}

\keywords{climate models $|$ calibration $|$ machine learning $|$ ...} 

\usepackage{tcolorbox}

\newcommand{\plref}[1]{page~\pageref{p-#1}, line~\lineref{l-#1}}
\newenvironment{answer}{\color{color2}}{}

\hypersetup{colorlinks=true,urlcolor=blue,citecolor=red}
\newcommand{\urlref}[2] {\href{#1}{#2}\footnote{\url{#1}, retrieved \today.}}

\newcommand{\order}{\ensuremath{\mathcal{O}}}
\newcommand{\state}{\ensuremath{\mathbf{x}}}
\newcommand{\param}{\ensuremath{\mathcal{M}}}

\title{Are General Circulation Models Obsolete?}

\begin{abstract}
  Traditional general circulation models, or GCMs -- i.e. 3D dynamical models with unresolved terms represented in equations with tunable parameters -- have been a mainstay of climate research for several decades, and some of the pioneering studies have recently been recognized by a Nobel prize in Physics. Yet, there is considerable debate around their continuing role in the future. Frequently mentioned as limitations of GCMs are the structural error and  uncertainty across models with different representations of unresolved scales; and the fact that the models are tuned to reproduce certain aspects of the observed Earth. We consider these shortcomings in the context of a future generation of models that may address these issues through substantially higher resolution and detail, or through the use of machine learning techniques to match them better to observations, theory, and process models. It is our contention that calibration, far from being a weakness of models, is an essential element in the simulation of complex systems, and contributes to our understanding of their inner workings. Models can be calibrated to reveal both fine-scale detail, or the global response to external perturbations. New methods enable us to articulate and improve the connections between the different levels of abstract representation of climate processes, and our understanding resides in an entire hierarchy of models where GCMs will continue to play a central role for the foreseeable future..
\end{abstract}

\dates{This manuscript was compiled on \today}
\doi{\url{www.pnas.org/cgi/doi/10.1073/pnas.XXXXXXXXXX}}

\begin{document}

\maketitle
\thispagestyle{firststyle}
\ifthenelse{\boolean{shortarticle}}{\ifthenelse{\boolean{singlecolumn}}{\abscontentformatted}{\abscontent}}{}

\section{Introduction}
\label{sec:intro}



The general circulation model, or GCM, is
a mainstay of research into the evolving state of the Earth system over a range of timescales. The term dates back to the very origin of numerical simulation of the atmosphere \cite[e.g.,][]{ref:phillips1956,ref:lorenz1967}. The equations governing the general circulation of fluids on a spinning sphere use the basic Navier-Stokes equations, whose form specialized for the planetary circulation were first formulated at the turn of the 20th century \cite[e.g.,][]{ref:bjerknes1904,ref:bjerknes1919}. However, closed-form solutions are not readily available, and their use as research and prediction tools had to await the advent of numerical solution in the 1950s \cite{ref:charneyetal1950}.

The 2021 Nobel prizes in Physics honour some of the work done with GCMs. The first formal global warning of anthropogenic climate change, the Charney Report \cite{ref:charneyetal1979}, was substantially based on the pioneering work of Syukuro Manabe, who confirmed 19th century speculations on the warming effect of adding CO$_2$. While normally attributed to Tyndall and Arrhenius, the earlier work of Eunice Foote has recently come to light \cite{ref:sorenson2011}. She in fact presciently remarked, \urlref{https://www.climate.gov/news-features/features/happy-200th-birthday-eunice-foote-hidden-climate-science-pioneer}{``An atmosphere of that gas would give to our earth a high temperature''}. While Foote and others were talking principally about the radiative effects of CO$_2$, it was Manabe and others who included dynamical considerations, the transport of heat both vertically through convection \cite{ref:manabewetherald1967}, as well as from the Equator polewards through atmospheric and oceanic circulation \cite[e.g.][]{ref:manabewetherald1975}.
Besides, GCMs also play a central role in the work of another of the 2021 winners, Klaus Hasselmann, who laid the groundwork for the statistical methods behind the field of detection and attribution of climate change \cite[e.g.,][]{ref:hasselmann1993}.
The detection of climate change requires extracting the signal of forced response in
simulations from natural variability, and the attribution of it to external climate forcing agents, such as CO$_2$ emissions, again requires counterfactual runs of a GCM where that particular forcing is absent.

It may seem an odd juncture, when a Nobel prize has just been awarded for GCM-based work, to speculate on the obsolescence of the GCM. However, there has been a considerable body of literature for a while, arguing that the limitations of GCMs require a major overhaul for further progress in climate modeling. It has been noted \cite[see e.g.,][]{ref:bonyetal2013}, that the bounds of uncertainty on equilibrium climate sensitivity (ECS: the asymptotic response of a model climate to a doubling of CO$_2$ concentration) has not significantly diminished since the Charney Report \cite{ref:charneyetal1979}. Furthermore, a systematic synthesis of multiple lines of evidence to constrain ECS in \cite{ref:sherwoodetal2020} indicates in several places a diminishing role for GCMs relative to other sources of information.
Some have taken a leap from here to assert that the entire project of parameterization -- the discovery of parsimonious representation through insight or mathematical methods -- may have no future, \citep[e.g.,][]{ref:bjornetal2020,ref:slingoetal2022}, and that large-scale computation is the way forward.

It is perhaps no accident that this debate takes place at a particular inflection point in the history of computing
\cite{ref:balaji2021}, where it is now possible to marshal and extract information from data at unprecedented scale, the era of big data and machine learning. These methods have led to some spectacular successes in various fields: AlphaFold for example can decipher the structure of complex molecules directly from data \cite{ref:jumperetal2021}. This has led to speculation that we might have entered the era of \urlref{https://www.theguardian.com/technology/2022/jan/09/are-we-witnessing-the-dawn-of-post-theory-science}{``post theory science''}. This is a fierce debate in many fields, whether large-scale structure emerges directly from the addition of detail and data, and where the limits of models built from data might lie. We explore this debate here in the context of the modeling of climate, a complex system undergoing slow but inexorable global changes, but where the details matter as well.

The debate poses questions that resonate across all fields of science that use large-scale data and computation as a pillar of the scientific method, alongside theory and observations. In the discussion at the end, Sec.~\ref{sec:hierarchy}, we will point out certain parallels, particularly with the debate in neuroscience surrounding the Human Brain Project. Leading up that discussion, the paper below will begin with an account of the structure of the GCM from the time of Manabe's pioneering studies to the present day, Sec.~\ref{sec:gcm}, followed by an analysis of some aspects of climate modeling which have exposed GCMs to criticism (Sec.~\ref{sec:ails}), interrogating the role of model resolution (Sec.~\ref{sec:resolution}), model calibration (Sec.~\ref{sec:tuning}), and the generation of counterfactuals (Sec.~\ref{sec:emulators}).
It is our assertion that parameterized GCMs can expect to get a new lease of life at this moment through sophisticated approaches to model calibration based on methods borrowed from machine learning. We begin with our account of the GCM.


\section{The structure of the GCM, from Manabe to present day}
\label{sec:gcm}

The general circulation of the atmosphere and ocean can be described by equations of fluid
flow and thermal energy transport and exchange. The numerical solution of a discretized
form of these coupled partial differential equations can be written in the form

\begin{equation}
\label{eq:state}
  \frac{\partial\state}{\partial t} = R(\state) +U(\state) + P(\state) + F
\end{equation}

where \state\ is a state vector consisting of mass, momentum and energy associated with fluid elements, as well as other quantities that can contribute to changes in state, which could include various phases of water in the atmosphere, salinity in the ocean, and trace elements,
CO$_2$, methane, dust, and other species, of natural or human origin. $R(\state)+U(\state)$ represents the dynamics, the Navier-Stokes equation. Because of the discretized nature of the numerical model, only part of the fluid motions are explicitly resolved ($R$). The unresolved ($U$) fluid motions (with scale smaller than one or several grid points) are represented in the form of a closure, \emph{i.e.,} a representation of subgridscale dynamics in terms of resolved-scale state variables. 
$P$ represents other processes that contribute to the thermodynamics: these can include diabatic processes associated with phase changes of water in the atmosphere, leading to the formation of clouds and rain, and the influence of solar radiation 
, and the equation of state of a complex fluid with many constituents. $U$ and $P$ are often collectively referred to as the \emph{physics}. 
Finally $F$ represents the terms that are considered external (not simulated by the model) influences on the system, known as \emph{forcings}: these can include solar radiation, volcanoes, anthropogenic emissions of CO$_2$, geothermal heating and other radiatively active quantities, or particulate aerosols that can play a role in cloud formation. Both atmosphere and ocean are extremely shallow compared to the Earth's radius $R$ ($H/R\sim 10^{-3}$, where $H$ is the fluid height) so that the numerical treatment of vertical dynamics is generally different from the horizontal..




In \cite{ref:manabewetherald1967}, Manabe
and Wetherald considered the time-asymptotic balance between the destabilization of an atmospheric column by radiation (warming at the surface, cooling aloft) and stabilization by convection, which transports heat vertically. Radiative-convective equilibrium (RCE) mediated by water vapor and CO$_2$ yields the basic behavior of global warming in a single atmospheric column. Later work,
\citep[e.g.,][]{ref:manabewetherald1975,ref:manabeetal1965} extended it to include horizontal transport of heat as well, from equator to the poles, confirming the single-column result. To give a sense of the computational size, a typical GCM resolution today is about 50~km or less, compared to 500~km in \cite{ref:manabewetherald1975}, and vertical resolutions have increased by a comparable factor: at this resolution, the required temporal resolution is measured in minutes. At these scales, the spatial grid may encompass $2\times 10^7$ points, and a simulation of 100~years in length requires $3\times 10^6$ timesteps. In the decades since the pioneering work of Manabe, considerable ingenuity has gone into creating fast and accurate solutions to the dynamics and increasing its resolution. 

\begin{figure}[htb]
  \begin{center}
    \includegraphics[width=\columnwidth]{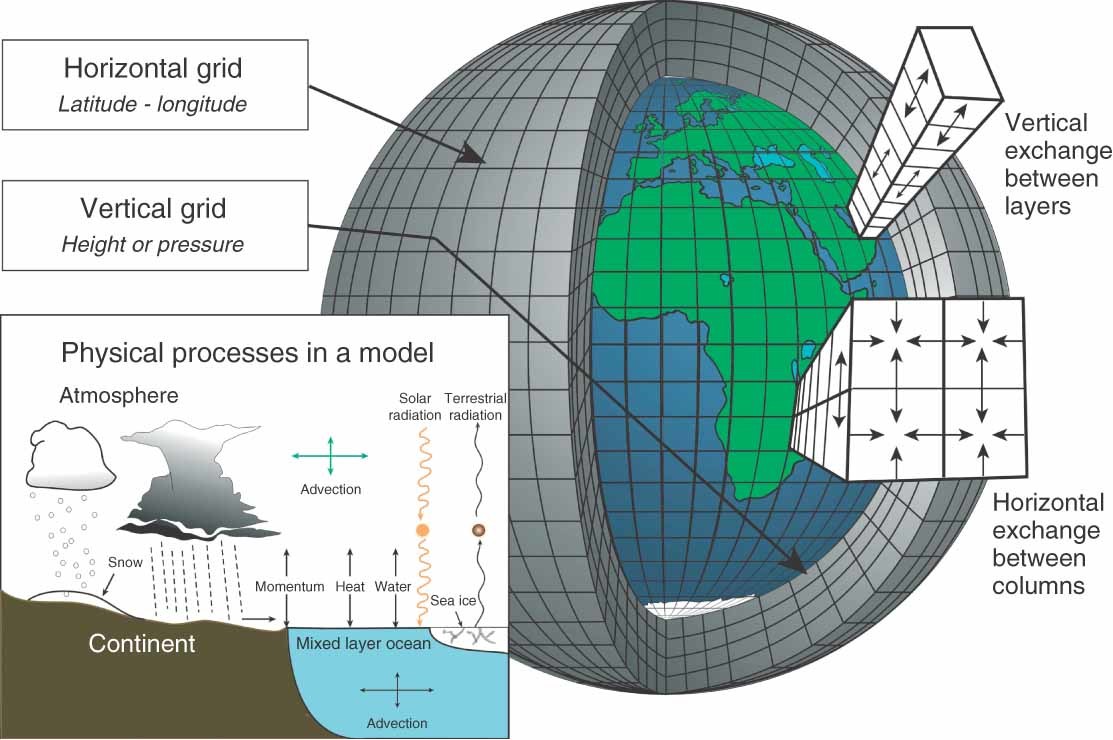}
  \end{center}
  \caption{Fig.~2 from \cite{ref:edwards2011}: the classic structure of a GCM. A similar column structure is used in the ocean
    as well. With permission of author.}
  \label{fig:gcm}
\end{figure}

An even more significant site of creativity has been in the physics ($U(\state)+P(\state)$) which now encompasses myriad processes in the atmosphere and ocean contributing to the transport of heat and other quantities: these include processes associated with clouds and precipitation, mixing by unresolved motions in the ocean, and an increasingly sophisticated treatment of the terrestrial and marine biosphere, with the contribution of trace elements and aerosols, whose natural and anthropogenic emissions also play a role in understanding the response of the climate system to changes in $F$. The structure of the GCM (Fig.~\ref{fig:gcm}) now consists of a number of \emph{parameterizations} representing individual processes and feedbacks:

\begin{equation}
  \label{eq:param}
  U(\state)+P(\state) =  \sum_{p} \param ( \state, \lambda_p)
\end{equation}

where $\lambda_p$ represents a set of parameters that may be empirically set through comparison with observations or theory. Typically these are \emph{bulk} quantities, representative values at the resolution of the model discretization. The climate is a multiscale system, encompassing processes from microscales (cloud droplet formation, stomatal exchange) to megascales (orbital changes, plate tectonics). Transport of heat and CO$_2$ into the ocean abyss occurs in the planetary scale on times measured in millennia.

We introduce a vocabulary for problems addressed with numerical models, where \state\ in Eqn.~\ref{eq:state} represents the state of the Earth system, and identifying sources of error and uncertainty in predictions from these models.


\begin{itemize}
\item Note first that Eq.~\ref{eq:state} represents the time trajectory of \state. The trajectory itself is the \emph{weather}. The system is chaotic \cite{ref:lorenz1963}, which represents the first source of uncertainty. Numerical weather prediction has improved over decades with better models, observations, and the techniques of \emph{data assimilation}, which constrain trajectories to stay close to observations in a least-squares sense \cite{ref:baueretal2015}.
\item The \emph{climate} is the set of preferred states of the system, its attractors, discovered by running trajectories for a long time, and averaging over the weather. If $F$ is held constant, the climate should be stationary, and the fluctuations around that state are its \emph{internal variability}.
\item The forcing $F$ is time-varying, with both natural and anthropogenic contributions, and the climate change problem consists in separating the forced response from internal variability. For purposes of policy related to the climate emergency, this constitute studying the response to \emph{scenarios} \cite[trajectories of $F$, see e.g.,][]{ref:tebaldietal2021}.
\item Abrupt transitions, or \emph{tipping points}, are changes in climate which are very fast relative to the rate of change of $F$. For instance, orbital changes are known to lead to long term changes in climate, such as the glaciation-deglaciation cycles. Yet, the climate record contains transitions such as Heinrich events \cite{ref:broeckeretal1992,ref:mcmanusetal2004}, a rapid collapse and resumption of the planetary scale meridional overturning circulation (MOC). We shall return several times to the MOC in this article as a canonical problem of concern for climate. The long intrinsic timescales of the forces shaping the MOC pose a unique set of
challenges. Studies of individual processes \cite[e.g.,][]{ref:schneideretal2019} also show that the system may be capable of abrupt (relative to the rate of change of $F$) transitions.
\item \emph{Parametric uncertainty} arises from imperfect constraints, either from observations or theory, on $\lambda_p$. \emph{Structural uncertainty} and structural error arise when no values of $\lambda_p$ can fit the known constraints, leading to the conclusion that the form of the equations in \param\
could be improved.
(It is of course possibly to formulate structural uncertainty as parametric, by having a parameter that chooses one equation structure over another!)
It has been noted that averaging over different GCMs (individual formulations of $R$, $U$ and $P$) can disguise structural error given statistical independence among GCMs \cite[e.g.,][]{ref:reichlerkim2008}. However, some biases remain common and persistent across many generations of models, suggesting gaps in understanding, or \emph{epistemic uncertainty}
\cite{ref:abramowitzetal2019}.
\end{itemize}


For the purposes of this Perspective, we define the GCM as the tool used to study the response of the climate system to changes in $F$. The tool is broadly similar (and in many cases is based on the same model code) to the models used for numerical weather prediction (NWP). In both cases, we follow trajectories of \state\ in time from a specified initial state. For weather the trajectory itself is the solution, while for climate we are interested in the attractors of the landscape where the trajectories lie. Besides chaotic uncertainty, we must also contend with the \emph{structural} and \emph{parametric} uncertainty associated with the physics terms $U$ and $P$. Finally $F$ and its time rate of change (usually ignored for weather, though not for climate) is also uncertain, as emissions trajectories are unpredictable \cite{ref:hawkinssutton2009}. For timescales of interest to address the climate emergency, including the possibility of abrupt transitions, we generally need to run simulations of at least \order(100)~simulated years (SY) in length. The exploration of the different sources of uncertainty will require sampling at least \order(100) model settings,
as we shall show below in Sec.~\ref{sec:emulators}. And finally simulations must complete in a reasonable time relative to the human attention span, which we define as \order(100)~days. A GCM is defined for the purpose of this study as \emph{an engine for simulating the Earth system, capable of running 100~instances of 100~SY}
in a reasonable amount of time, and 
given adequate computing resource. The same tool may be configured for other purposes (e.g process studies on a limited domain or in highly idealized settings, or higher resolution but shorter duration), but these constitute the minimum requirements for studying the response to changes in $F$.


\section{What ails the GCM?}
\label{sec:ails}


 




There are several diagnoses of the weaknesses of GCMs. There is first the argument that the column abstraction breaks down in the presence of large-scale organization: the mesoscale organization of cloud systems in the atmosphere \cite{ref:schumacherrasmussen2020}. A similar argument can be made for ocean dynamics as well, where mesoscale turbulence in the form of persistent eddies is able to deposit energy and momentum away from the source \cite[e.g.,][]{ref:rhines1979}, in the form of Agulhas Rings for example \cite{ref:jacksonetal2019}. Such non-local effects of unresolved terms call into question the structure of the GCM that has been used since
Manabe's pioneering calculations. Some aspects of the climate system have resisted efforts at representation in models, with stubborn biases and uncertainties. This has led some to question whether such processes are parameterizable at all
\cite[e.g.,][]{ref:slingoetal2022}. Following this line of thinking, it is now often contended that nothing short of resolving finer-scale motions, coupled with assimilation of present-day observations to control model biases, will in fact address these shortcomings, and that a future generation of models
will address these issues through substantially higher resolution.

A second criticism of GCMs is around the consideration of the ``tuning'', or calibration, of climate models. As noted above, unresolved physics ($U+P$) is represented using equations with parameters constrained within some range by observations or theory. The coupled system is then further subjected to global constraints such as top-of-atmosphere energy balance \cite{ref:hourdinetal2017}. The fact that the models are tuned to reproduce some features of the observed planet is in some quarters viewed as rendering the results suspect.

Finally, GCMs are now numerous (\urlref{http://esgf-ui.cmcc.it/esgf-dashboard-ui/data-archiveCMIP6.html}{114 models from 44 institutions at time of counting}). Viewed as an ensemble of simulations, they embody considerable structural or epistemic uncertainty, in the form of differing representations of the unresolved scales. The GCMs are not all statistically, or in terms of model code, independent of each other \cite{ref:abramowitzetal2019}, and different evaluation metrics yield different, and contradictory measures of model quality.
The uncertainty bounds have, if anything, increased between the last two climate model assessment cycles, and the recently published 6th Assessment Report \cite{ref:masson-delmotteetal2021} notes that many models now produce ECS
values outside the assessed ``very likely'' range, leading to an enhanced role for emulators
\citep[e.g.,][]{ref:hausfatheretal2022}, as discussed in Sec.~\ref{sec:emulators}. Models used in service of decision-making and policy, including
those used in the recently concluded IPCC AR6, rely on emulators allowing of rapid exploration of multiple future scenarios or through the use of statistical techniques for the correction of biases. Furthermore, it is increasingly noted that models pegged to present-day climate do not do a good job of representing the climate fluctuations of the past, including past warm climates that may hold lessons for the climate emergency
\citep[e.g.,][]{ref:valdes2011}. We address these concerns in turn.



\section{What resolution is enough?}
\label{sec:resolution}

We begin with the question of resolution. As noted above, one suggested remedy for the weakness of GCMs is to increase the resolution until some of the non-local phenomena alluded to above are in fact resolved \cite[for instance]{ref:scharetal2020}. The computational expense of such a model would require a substantial boost to computing capability \cite{ref:neumannetal2019}.




Atmospheric and ocean dynamics fit within the broad contours of geophysical turbulence. At very large (planetary) scales, this looks like 2-dimensional turbulence,
known to have an energy spectrum with a $k^{-3}$ spectrum. At smaller scales, baroclinicity starts to play a role, and 3-dimensional turbulence with a $k^{-5/3}$ spectrum. This is in fact observed in the atmosphere, seen for example in \cite{ref:nastromgage1985}. Similar spectra are observed in ocean turbulence as well \cite{ref:stammerboning1992,ref:capetetal2008}. The key feature to underline for this discussion is that the 3-dimensional energy cascade continues all the way down to molecular scales. There is no fundamental scale separation in turbulence. Any truncation applied in order to create a discrete representation for numerical purposes is an arbitrary one, usually
constrained by the available computing power.


We can look beyond turbulent energy spectra to the specificities of certain dynamical phenomena. In the atmosphere, one of the key phenomena of interest is moist convection. Since the pioneering work of Rayleigh and others shows that fluids heated from below will overturn, with the overturning motion in the form of ``cells'' that roughly scale with the height of the convecting fluid. In the case of atmospheric convection, this includes deep convection, which roughly scales with the height of the tropopause \cite[$\sim$10~km, see e.g.,][]{ref:balajiredelsperger1996}, such as thunderstorms or tropical cyclones, and shallow convection within the planetary boundary layer (depth $\sim$1~km), which can take many forms, including large stratocumulus decks 1000s of km in extent, playing a significant role in the planetary albedo and heat balance. GCMs have traditionally tried to represent these in parameterizations of the vertical transport of momentum, heat and moisture, as well as other tracers, by convection.
Subgrid closures of deep convective processes can be based on an assumption of quasi-equilibrium between synoptic-scale destabilization of the column and the stabilization by convection \cite[e.g.,][and its descendants]{ref:arakawaschubert1974}. The slow rate of advances in these methods \cite{ref:randalletal2003} have led to efforts where parameterizations are replaced with embedded cloud-resolving models (CRMs), known as ``super-parameterization'' \cite{ref:khairoutdinovrandall2001}, or even further to replace shallow convection as well with large-eddy simulation (LES) models \cite{ref:parishanietal2017}. It is not clear if since their inception this class of models have justified their extreme computational expense in terms of improved climate simulation.

Many aspects of clouds, such as the representation of moist convection requires us to invoke microphysical processes involved in the condensation of water vapour and the formation of falling hydrometeors. But the motions themselves can be captured in non-hydrostatic models. Limited area modeling of deep convection dates back to the 1970s, and advances in computing capacity in the intervening decades makes global
cloud-resolving models (GCRMs, also known as global convection-permitting models, global storm-resolvind models, etc.)) within reach \cite[see e.g.,][for a comparison across multiple GCRMs, the DYAMOND experiment]{ref:stevensetal2019}. These models typically have horizontal resolutions in the range of 1--5~km, considered sufficient to capture mesoscale convective organization, and at least marginally resolve individual convective events.
Aspects of convective organization, such as the formation of gust fronts with downstream surface density currents
\cite{ref:grandpeixlafore2010}, remain below the resolution of GCRMs, and must be parameterized, as they are responsible for the initiation of new convective events. Boundary layer convection is responsible for shallow clouds, and requires at least 10X higher resolution: this is in fact one of the largest sources of uncertainty in the current generation of models \cite{ref:zelinkaetal2020}. This class of clouds will not be resolved by km-scale models. And of course, water as vapour or condensate is radiatively active, and particulates play a role in cloud formation as well. Surface exchanges at very fine scale mediate aerosol emissions \cite[e.g.,][]{ref:leeuwetal2011}. Many of these aerosol-radiation-cloud (ARC) processes take place at micron-scale and will be forever outside any conceivable resolution of a numerical model
on any known computational technology in the literature today.

\begin{figure}[htb]
  \begin{center}
    \includegraphics[width=0.49\columnwidth]{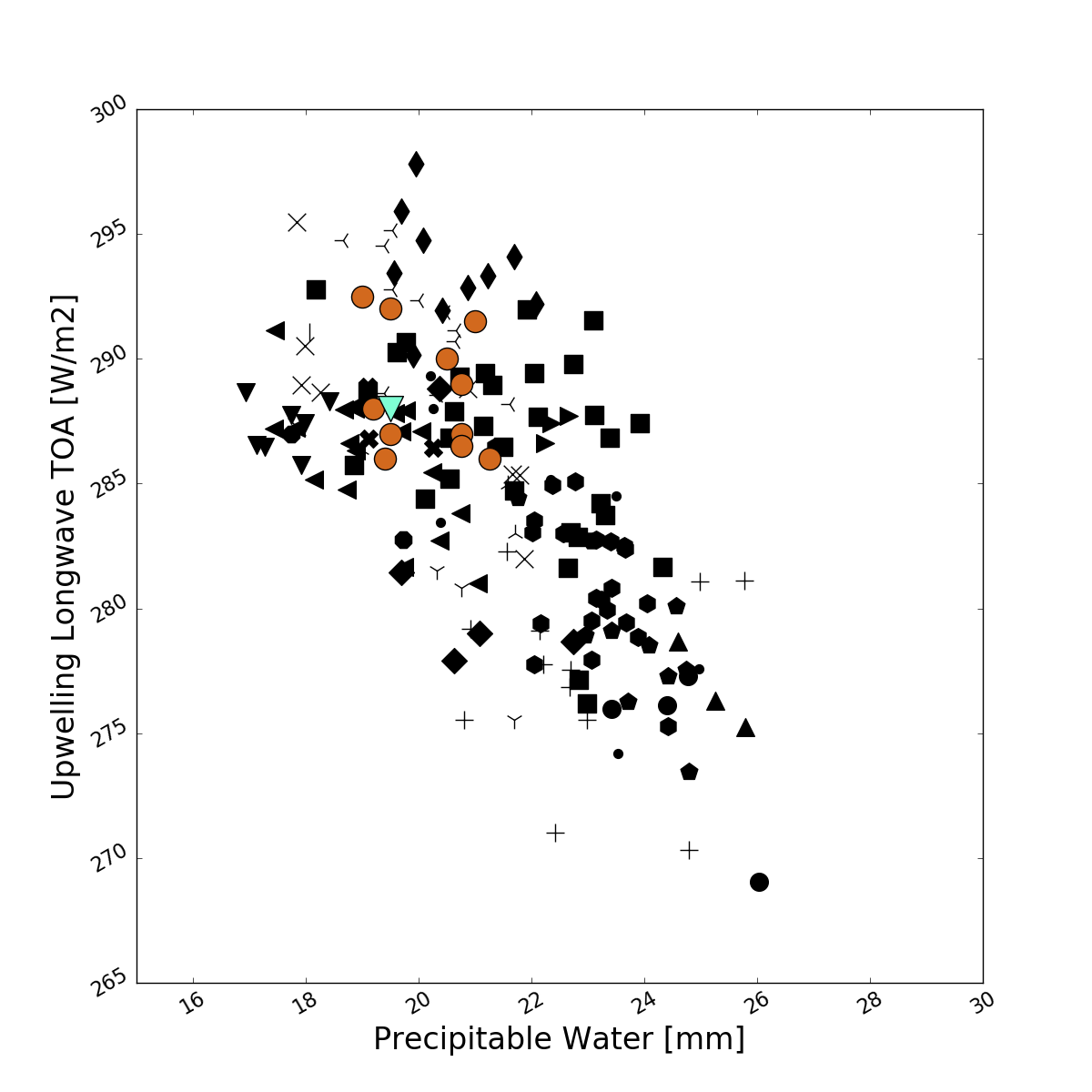}
    \includegraphics[width=0.49\columnwidth]{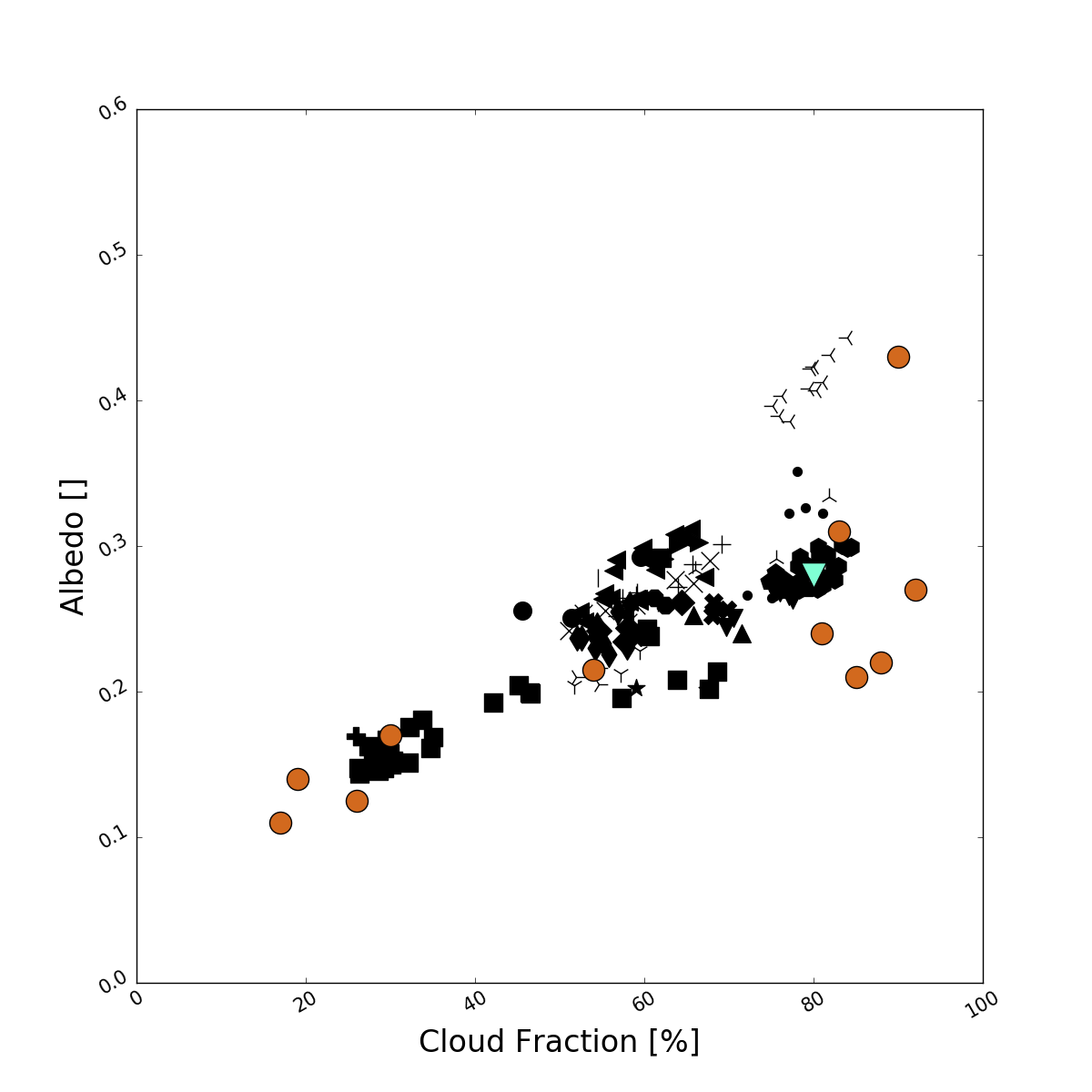}
  \end{center}
  \caption{Comparison of observable properties of cloud fields (August monthly-mean outgoing longwave radiation, precipitable water, albedo, and cloud fraction) averaged over the Southern Atlantic (15W-10E;18S-5S) in observations (green triangle), DYAMOND GCRMs \cite[orange circles, from][]{ref:heimetal2021} and in CMIP6 GCMs. The added resolution in GCRMs reduces structural uncertainty in some respects but not others.}
  \label{fig:dyamond-cmip6}
\end{figure}

CRMs (and their global incarnation, GCRMs), and LES models of the turbulent and cloudy boundary layer are widely used to study processes that cannot be resolved in GCMs, but perhaps we can learn from them to inform the development of parameterizations. Such studies are usually mediated by single-column models \cite[SCMs, see e.g.,][]{ref:randalletal1996}.
Such studies often show that the disparity between different CRM formulations remains comparable to those between GCMs. In a set of RCE \cite[the same problem treated in a column by][]{ref:manabewetherald1967} comparisons across both CRMs and GCMs \cite{ref:wingetal2020}, the uncertainty spread across CRMs and GCMs with parameterized convection were quite comparable \cite{ref:beckerwing2020}. In the first extensive comparison of GCRMs \cite[necessarily short runs, only 40~days compared to the GCM timescales outlined in Sec.~\ref{sec:gcm}][]{ref:stevensetal2019}, there was considerable inter-model variability \cite{ref:heimetal2021}.
In Fig.~\ref{fig:dyamond-cmip6} we show a comparison of the relationship between precipitable water and outgoing top-of-atmosphere longwave radiation and albedo and low cloud cover from \cite{ref:heimetal2021} with the same quantities from the CMIP6 experiment. While the CRMs are closer to observations for the precipitable water and outgoing LW radiation, the spread in low cloud cover and albedo is just as wide in GCRMs as in GCMs.
The model spread in DYAMOND has been traced to differences in the treatment of boundary layer convection \cite{ref:christensendriver2021}.


LES structural uncertainty tells a similar story: comparisons across LES models of the boundary layer with or without clouds still shows
some reduction in spread in certain cases \cite{ref:siebesmaetal2003}, but considerable
spread in response to changes in configuration in specific aspects such as microphysics \cite{ref:zantenetal2011}, turbulence closures \cite{ref:beareetal2006,ref:couvreuxetal2020} and numerics \cite{ref:presseletal2017}.

Similarly, the representation of oceanic fronts, eddies and currents is markedly different depending on the type of model that is used to represent them, whether turbulent mixing by eddies is resolved or parameterized. The ocean depth varies widely between pelagic zones and the abyss, and consequently the Rossby deformation radius -- the length scale representing when mesoscale eddy mixing is significant -- varies from \order(100)~km in the Equatorial open ocean, to \order(1)~km or less near coastlines and toward the poles \cite{ref:hallberg2013}. Attempts to build ``scale-aware'' parameterization of eddy mixing \cite[e.g.,][]{ref:jansenetal2015,ref:bachman2019}, must contend with these variations.






Other features such as vegetation, not discussed in this article, also exhibit heterogeneity at any conceivable resolution. While there is no doubt that in general increasing resolution reduces the number of processes needing to be parameterized
(although placing increasing resolution demands on observations as well), and in many cases increases the fidelity of simulations particularly in the short term, in the \emph{climate} context this must be weighed against the expense, which limits the simulation duration and consequently ability of models to realize the aspects that regulate the climate on long timescales.
The absence of any target resolution where one could argue that everything of interest to climate is resolved, implies that these tradeoffs are always with us.

\section{Are there ``untuned'' models?}

\label{sec:tuning}


The question of the ``tuning'', or calibration, of GCMs, has been a fact of life since their inception, often not clearly described in the literature,
until recent efforts to document the role of tuning in model development \cite{ref:hourdinetal2017,ref:schmidtetal2017}.
On the scientific level, these attempts highlight the central role of tuning in modeling, and open up new avenues in the use of
automatic calibration techniques from machine learning.

In broad terms, we can
define the process of tuning as one of finding suitable values of $\lambda_p$ in Eq.~\ref{eq:param} that best fit observations or theory, identifying it as an intrinsic and universal aspect of model development.
We seek to achieve both fidelity to each process \param\ in the model, as well as respecting global constraints across the coupled system: for example, conserving energy and mass, including of individual agents in the climate system, such as water. The global constraints must be applied when coupling models at any resolution. Tuning is thus a multi-step process, where individual parameterizations \param\ are first tuned separately to within a desired tolerance, but which may then be refined in a second stage after the coupled model is built. Thus tuning, which is often seen as an optimization of a loss function, may be redefined as identifying the subspace of parameters compatible with a number of constraints. While the procedures can be onerous, the process of calibration is central to model development and the way teams learn how parts of the coupled system respond to changes in others. The coupled system can yield surprises: in one example \cite[the NOAA/GFDL model][]{ref:heldetal2019,ref:zhaoetal2018b}, the coupled system had an ECS higher than was expected during the development of individual components.
Thus, \emph{model calibration is not a weakness of models, it in fact holds the key to how model developers learn how their model behaves, and consequently how the Earth system regulates itself.}

Historically, the tuning of models
has been found to be expensive, if one uses the GCM itself as the forward model. In particular, key circulation features such as the MOC may be sensitive to tuning in ways that reveal themselves after simulations of \order(100)~SY \cite{ref:heldetal2019}. Tuning such models ``by hand'' can be an inefficient scattershot exploration of a small amount of the possible space of parametric uncertainty. This is perhaps one reason why some observers hold tuning in low regard.
There is also the risk of tuning to a state containing compensating errors, that satisfy the constraints but for the wrong reasons.


As noted earlier, computing technology at the present time favors the algorithms of machine learning (ML), the ability to emulate or extract patterns seen in large datasets. Tuning is a constant concern in the construction and optimization of
methods, such as choosing the width, depth and structure of a neural net:
aspects often referred to as \emph{hyperparameters}, as they are of the network, not of the process being emulated). They are chosen to meet the requirements of fidelity against the training dataset. Some data is often withheld to guard against overfitting to noisy data, and the withheld (``out of sample'') data can be used to validate the result as generalizable to novel situations not seen during training. ``Physics-informed'' ML \cite{ref:karniadakisetal2021}, where the loss function can be made to penalize violations of global constraints such as conservation laws, also has a parallel to the tuning process described above.

\begin{figure}[htb]
  \begin{center}
    \includegraphics[width=\columnwidth]{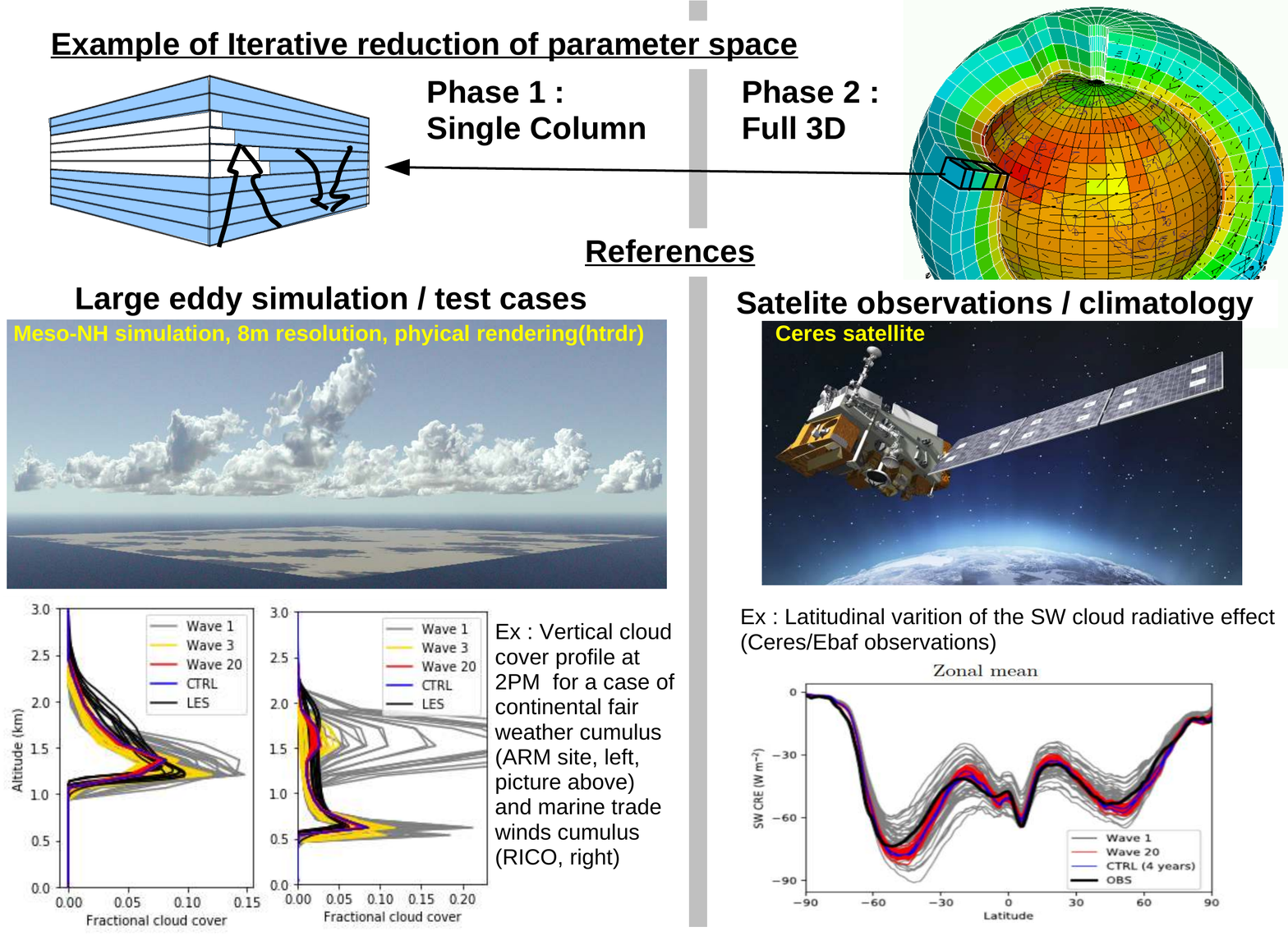}
  \end{center}
  \caption{A schematic of the use of LES models for parameter estimation, adapted from
    \cite{ref:couvreuxetal2021}. LES models of different observed cloud regimes are used to build emulators to estimate parameters governing the behavior of column-averaged quantities. Successive waves of history-matching deliver the minimal parameter space consistent with the LES data, the NROY space. The effect of clouds on absorbed shortwave radiation is reduced by a comparable amount to the previous laborious tuning ``by hand'' using the same target.}
  \label{fig:hightune}
\end{figure}

In the HighTune project, an atmospheric boundary layer convection scheme based on the eddy-diffusivity mass-flux (EDMF) approach \cite{ref:hourdinetal2002,ref:rioetal2010}.
is calibrated to match results from an LES simulation \cite{ref:couvreuxetal2021,ref:hourdinetal2021}. The EDMF scheme, which has both upgradient (organizing) and downgradient (dissipative) components, is
structured for problems such as cloudy convection and turbulence, but has parameters that must be empirically determined in a variety of boundary layer regimes. The process is illustrated in Fig.~\ref{fig:hightune}. A variety of cloud regimes is simulated in an LES constrained by observations: the LES serves as the ``truth'' for tuning. A single-column representation of EDMF is then compared against LES output using the ``history-matching'' (HM) method of 
\cite{ref:williamsonetal2013}, exploring the range of uncertainty simultaneously across multiple parameters at number of points. The full range of parametric uncertainty is then explored using emulators, as described in \cite{ref:couvreuxetal2021}. 
Rather than seeking a single optimum $\lambda_p$, HM seeks only remove \emph{implausible} regions of parametric space from consideration, leaving a \emph{not ruled out yet} (NROY) region for consideration. Any parameter values within NROY are possible valid values, and
new constraints can be progressively added as needed. Finally, the HM method can also serve as a means of diagnosing structural error: a null NROY space (i.e no possible values of $\lambda_p$ permit \param\ to meet the desired constraints within the chosen tolerance) indicates that the representation in \param\ needs to be refined.

An independent effort from the \urlref{https://clima.caltech.edu/}{CliMA} project uses a somewhat different approach, but with the same ends, namely to calibrate an EDMF scheme with a set of parameters and their uncertainty bounds \cite{ref:dunbaretal2020,ref:clearyetal2021}. There are also efforts to build a ``library'' of cloud regimes \cite{ref:shenetal2021} for use in training. As noted above in Sec.~\ref{sec:resolution}, LES models carry their own structural uncertainty. The \urlref{https://www.umr-cnrm.fr/spip.php?article930&lang=en}{DEPHY} project
seeks to build systematic benchmarks and training datasets for similar efforts elsewhere.

This represents the first step toward a more systematic and objective exploration of parametric uncertainty than is possible when GCMs are directly used as the forward model. These methods are now being extended to consider the tuning of slower processes in the ocean, following \cite{ref:williamsonetal2017}. Some of the subsequent problems of tuning coupled systems with multiple timescales are still being explored using highly simplified models \cite{ref:lguensatetal2022}. Preliminary results indicate that compensating errors and other defects inherent in the tuning of coupled systems may still be present, but it may be possible to diagnose those within the HM method. 

\section{Why do we need counterfactual Earths, and where do they come from?}
\label{sec:emulators}

In Sections \ref{sec:resolution} and \ref{sec:tuning}, we have outlined a procedure for creating a tool for numerically simulating a very complex multiscale system at a maximal level of detail given computing limits, and calibrating the system within those limits to resemble the observed Earth as closely as possible. We use this system to study the Earth's climate and its history, the appearance of life and the maintenance of an atmosphere, ocean, and land surface suitable for life, the fluctuations of the past and the recent period of relative stability that allowed for the possibility of settlements and agriculture, and the consequences of the Industrial Revolution leading to the current climate emergency. As an object of science, the Earth poses a key problem, in that there is only one instance of it, and one temporal trajectory of its state, that we can observe. Simulated Earths remain our only means of exploring different hypotheses about how the system works, other trajectories it might have taken, and might take in the future. It also remains our only means of exploring responses to the climate emergency, and understanding and predicting the impacts of different policy choices of mitigation, adaptation, and resilience-building.

Using techniques pioneered by Hasselmann, another of 2021's Nobels in Physics \cite[e.g.,][]{ref:hasselmann1993}, the techniques of \emph{detection and attribution} have helped understand the role of different forcing agents in $F$, and their contribution to a changing climate. First, a comparison of present-day climate with a counterfactual simulation known as the ``pre-Industrial control'' (where $F$ is held constant at its value in 1850~CE), allows us to unequivocally detect that the climate has changed, beyond the bounds of
simulated random internal variability. Attribution of climate change to particular forcing agents is performed using either single-forcing runs, where all contributors to $F$ are held constant save one, or using agents grouped into \emph{greenhouse gases only}, \emph{natural only}, etc \cite{ref:gillettetal2016}, to tease apart the contribution of different forcings to climate change, as well as their non-linearities (as the contributions may not be simply additive). The number of individual forcings considered now number over 10 \cite[see, e.g., Fig.~2c in the IPCC AR6 Summary for Policymakers][]{ref:masson-delmotteetal2021}.
In addition, \cite{ref:tebaldietal2021} consider at least 8 pathways of future evolution of forcing. Together, such simulations represent a requirement for GCMs to explore hundreds of counterfactual pathways of climate evolution.

We draw attention to two major implications of this for the use of GCMs. First, GCMs are used to explore a counterfactual space, not directly constrained by observations. This includes chaotic uncertainty (different trajectories of \state\ under
changes in initial conditions, which provide a measure of the intrinsic internal variability), and to counterfactual settings of $F$, where various actual observed forcings are turned on or off. An individual GCM (a particular formulation of the terms $R$, $U$ and $P$) is first calibrated against theory and observations as described in Sec.~\ref{sec:tuning}, and once this is satisfactory, is used to explore an even vaster counterfactual space \cite[see e.g.,Fig.~4 from][]{ref:hawkinssutton2009}. The requirement to be able to simulate counterfactuals must be taken into account in the context of ML as well, as we shall discuss below in Sec.~\ref{sec:hierarchy}.

Second, in view of the computational expense of GCMs outlined in Sec.~\ref{sec:gcm} \cite[see also][]{ref:balajietal2016b}, it has proved prohibitively expensive to explore all the forcings, and their potential future pathways, using GCMs. The IPCC has turned instead to an extended use of \emph{emulators}. Note that unlike the emulators in Sec.~\ref{sec:tuning}, which attempt to mimic individual climate processes, these are emulators of the whole climate system, which attempt to predict the response of the entire system, usually in the form of an integrated measure such as the global mean surface temperature (GMST), to changes in $F$. Regional patterns of climate change can be inferred by coupling with techniques such as \emph{pattern scaling}
\cite{ref:tebaldiarblaster2014}. The emulators are all reduced complexity models of various flavors, ranging from relatively simple regressions trained on recent historical data, to dynamical systems models such as impulse-response models, to highly simplified, and usually 1D, GCMs \cite{ref:nichollsetal2020}. Their advantage is that they are typically millions of times faster than GCMs \cite{ref:jacksonforster2021}, although their lack of internal physical consistency poses epistemic risk.

One of the key measures of the climate response to CO$_2$, the ECS, is itself a counterfactual, as it
is based on an asymptotic equilibrium that is never observed in nature. It is nonetheless useful as a measure in order to project a range of responses to scenarios (possible future trajectories of $F$). Furthermore, while the recent (and most precise) observational record of the satellite era is too short to constrain GCMs adequately, there are other indirect means of placing limits on ECS, including paleoclimate data and constraints on individual processes contributing to the ECS \cite{ref:sherwoodetal2020}. The recent IPCC concluded that many GCMs were providing ECS outside the ``very likely'' range, and used emulators where ECS is a tunable parameter, to refine the consensus projections and their uncertainty bounds \cite{ref:jacksonforster2021}.

Reduced complexity models have also been used extensively to study the potential for abrupt transitions in the climate system, for which there is some evidence in the paleoclimate record \cite{ref:brovkinetal2021}. The MOC is a canonical feature of the climate system with the potential for metastability: under sufficient fresh water input in the North Atlantic, from retreating continental ice sheets for instance, the MOC can ``collapse''. Yet, GCMs exhibit metastability of the MOC less readily than reduced-complexity models, and eddy-resolving models even less so \cite{ref:gent2018}. A concern often expressed is that GCMs are too stable to perturbations \cite{ref:valdes2011}.

The ability to run a wide variety of century or millennial scale
simulations is essential for an evaluation of
many aspects of climate risk,
tipping points. High-resolution model trajectories constrained by assimilation of recent observations will do little to mitigate this concern. The discussion above in Sec.~\ref{sec:tuning} points to ways forward, objective methods of emulation of GCMs, just as GCMs themselves learn to emulate CRMs and LES models. This would alleviate the epistemic risk associated with emulators.

\section{What might a future modelling landscape look like?}
\label{sec:hierarchy}



In a celebrated 1972 essay \emph{More is different} \cite{ref:anderson1972}, Philip Anderson, another Nobel laureate in Physics, argued that many of our sciences abstract reality at different levels of complexity, and struck a cautionary note on the limits of assuming that one level of explanation is ``nothing more'' than an expression of aggregate behavior of the elements at a deeper level of abstraction. The understanding of such emergent behavior of complex networked systems (CNS) happens at multiple levels, and the reduction of the properties at one level to the one below may not be computationally tractable or of practical use, even if it is true in principle. The Earth system is an exemplar of a CNS, as discussed here: bringing together domains as different as fluid dynamics, radiative transfer, chemistry, biology and ecology, over a range of time and space scales. Yet the assembly as a whole appears to persist in stable states for millennia. The scientific puzzles related to the climate emergency center on understanding the emergent balance of terms perpetuating stable states, and what thresholds of these balances we are exceeding with anthropogenic perturbations.




As simulation is now a pillar of the scientific method,
one temptation at this point is to turn over the problem of emergence to the computer itself. Many diverse fields of science have attempted to simulate a CNS at a high level of network fidelity, explicitly incorporating all the interactions and feedbacks. The question is then, can such a numerical system spontaneously exhibit higher levels of organization? In neuroscience for example, we can pose this question in the context of the emergence of higher level brain function from the details expressed or simulated at the level of individual neurons. The Human Brain Project
\citep[HBP,][]{ref:amuntsetal2022}, attempts to do just this, and others have questioned the limits of this approach \cite[e.g.,][in the context of vision]{ref:fregnac2017}, or looking at ``top-down'' effects where the large-scale structure regulates the behavior at the neuronal level \cite{ref:ellis2019}.


The equivalent question to pose in climate science is whether the stable states of the climate, where \state\ from Eq.~\ref{eq:state} remains near an attractor despite fluctuations around it, emerge directly from the assembly of the system with all its details. We also would like to see if such a detailed simulation can accurately capture the responses of the system to changes in $F$. Despite the ``fast'' physics of the atmosphere and planetary surface, we have seen that the system contains natural and forced variability at ``slow'' timescales, regulated by such emergent features of the general circulation such as the Meridional Overturning Circulation (MOC). The timescales of changes in forcing, such as anthropogenic emissions since the Industrial Revolution, and of responses in features such as MOC demand modeling tools capable of century-long simulations under many possible forcing pathways, 100 simulations of 100~SY each, as argued in Sec.~\ref{sec:gcm}. We have reviewed the arguments for km-scale models in Sec.~\ref{sec:ails}, which are quite limited in capability compared to the GCMs
\cite[optimistically, perhaps capable of \order(10) simulations each of \order(10)~SY in length on the largest available computing, an order of magnitude below that,][]{ref:neumannetal2019,ref:wedietal2020}. Simultaneously, those exploring policy responses require very fast models for exploring many policy options under many forcing scenarios.



The sense that GCMs may be
obsolete comes from these conflicting demands: very high resolution for some key processes, which restrict the ability to explore and quantify uncertainties and study low-frequency variability; the need to explore many counterfactual scenarios for constructing climate policy, which cannot be guided simply by present-day observations; the long simulation times needed to understand prior episodes of abrupt climate change.

As noted in the comparison with neuroscience, this is a debate across many fields, on how much reliance to place on the most complex models capturing detail using the elements of the finest representation of all the elements in a complex system. Such simulations are sometimes now called ``digital twins'', borrowing a term from manufacturing \cite[e.g.,][]{ref:kritzingeretal2018}, where it originally meant an exact digital copy of an engineered system and its specifications. It has been increasingly used to describe complex systems, including living systems \cite{ref:wenetal2022}. But as noted in \cite{ref:wenetal2022}, for sufficiently complex systems, we need simulations at multiple levels of abstraction and complexity.

The MOC is a case in point. At the broadest level, the Earth is warmed by radiation from the Sun around the Tropics, and loses heat near the Poles, implying an Equator-to-Pole transport of heat in the ocean-atmosphere system. In the ocean, for the current planetary topography, this results in what is referred to as the ocean's ``conveyor belt'' for heat and energy \cite{ref:broecker1991}. Looking a little closer, we see important features such as the horizontal gyres that also contribute to the meridional transport of heat northward in the North Atlantic \cite{ref:lumpkinspeer2003}, an important feature regulating the climate on the continents around \cite{ref:palter2015}. Zooming further in, we can see that turbulent eddies continue to transport heat across the meridional flow in the North Atlantic \cite{ref:treguieretal2012}. All levels of explanation are broadly consistent with data, and while each rung of the ladder of complexity can be described in terms of residuals of a fine-scale balance from a level below, the balance is regulated by global constraints such as the Equator-to-Pole radiative imbalance. In short, \emph{it is the entire hierarchy that constitutes our understanding of the MOC}.

\begin{figure}[htb]
  \begin{center}
    \includegraphics[width=\columnwidth]{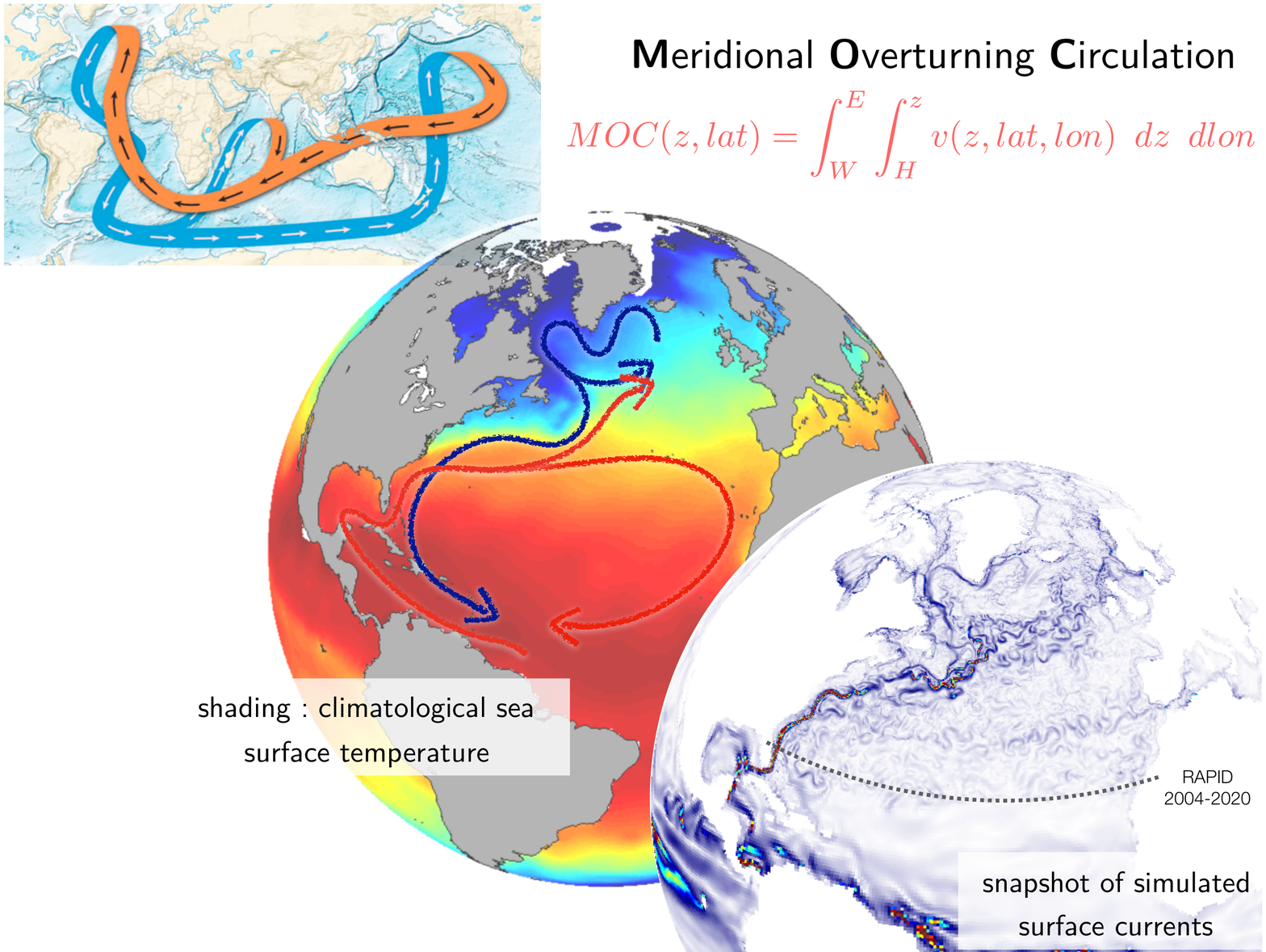}
  \end{center}
  \caption{Different views of the MOC, from models at various levels of Charney's ladder. They all exhibit fidelity to aspects of our knowledge of the MOC, and our understanding of the MOC lies in the composite.}
  \label{fig:hierarchy}
\end{figure}

In the modeling of complex multiscale systems, calling one of these layers a ``twin'' at the expense of others appears to be a rhetorical overreach \cite{ref:everssalles2021}. Indeed in speaking of the HBP in
\cite{ref:amuntsetal2022b}, the ambition is sharply circumscribed:

\begin{quote}
In constructing a ‘digital twin’ of a living organ, one is confronted by important challenges over and above those encountered when constructing the digital twin of an inanimate object. Therefore, the concept of the ‘digital twin’ in this context needs to be carefully defined to provide clarity on its limitations and to avoid creating unrealistic expectations of exact fidelity ... The digital twin is thus a copy in the practical sense, usually associated with a model of a function or process, and its power lies in its usefulness in dealing with relevant problems faced by its physical counterpart without the need (and certainly not the claim) of capturing every single detail thereof.
\end{quote}

In the climate context, this could describe any numerical model dating back to \cite{ref:charneyetal1950}!

We have demonstrated (Sec.~\ref{sec:tuning}) that there are methods for making the model hierarchy \emph{traceable}, by showing how to derive a parameterization on the basis of a model further up what Charney called the model ``ladder'' \cite{ref:balaji2021}.
It might be worth including such high-resolution benchmarks for the training of parameterizations in future model intercomparison projects for a new generation of GCMs calibrated using the methods outlined here.
Similar methods can be imagined to link the reduced complexity models of \cite{ref:nichollsetal2020} to GCMs, from which we can carefully explore how to remove the biases in the GCMs themselves, such as those outlined in Sec.~\ref{sec:ails}.

We contend in this article that the GCM remains the indispensable meeting point of these divergent directions. Its structure encapsulates our fundamental understanding of how the climate works, and represents an astute assembly of choices and tradeoffs that are versatile enough to meet the challenges outlined here. 
The GCM's column structure reflects the importance of separating the vertical in the model topology, and the importance of convection and its timescales in atmosphere and ocean. The structural independence of columns has been seen as a limitation, but new methods can use non-local predictors \cite{ref:wangetal2022}. The stochastic parameterizations mentioned in Sec.~\ref{sec:resolution} also impose non-local (in space and time) coherence to the stochasticity \cite{ref:leutbecheretal2017}.

Just as data assimilation techniques for constraining trajectories toward a time series of observations led to major improvements in NWP \cite{ref:baueretal2015}, these new methods, based on data from process models such as CRMs, or observations, as well imposing physical constraints, hold out the possibility of efficiently approaching the attractors of the system, the key feature distinguishing the climate problem from weather, as outlined in Sec.~\ref{sec:gcm}. The recommendation is to adopt a rigorous, transparent, and reproducible tuning process, rather than assuming tuning will simply disappear when we simulate at a high enough level of detail.

Finally, we do not wish to forget the use of models as pedagogical tools, for students to explore climate response, to have ``fun'' -- a point repeatedly made by Manabe after receiving the Nobel Prize, e.g in his first \urlref{https://www.youtube.com/watch?v=BUtzK41Qpsw}{Manabe Nobel press conference}. Models must in addition be easy to use to explore their sensitivity to counterfactual changes, explore novel and risky ideas on how it responds to perturbations.

The future modeling landscape must rest on the principle of a \emph{traceable model hierarchy} \cite{ref:balaji2021}.
Models at every level of the hierarchy have their own forms of structural uncertainty, as noted above: this uncertainty does not vanish at any conceivable model resolution possible on any known computational technology. Each model can be put to multiple uses and subject to diverse physical and computational constraints. The traditional GCM, with its ability to combine resolved dynamics with unresolved physics for a non-stationary Earth system, remains the crossroads between models built for other purposes: models that can resolve some of the physical uncertainties, but in limited settings, models that can be used to study transitions in the climate system that are abrupt events between millennia-long stable states, and emulators that produce corrected data for downstream users. Each of these involve tradeoffs sacrificing accuracy in one part of the climate system against another. The GCM will remain the essential element ensuring that these tradeoffs remain within reasonable limits for the entire Earth system.




\section*{Acknowledgments}
\label{sec:ack}

VB is supported by the Cooperative Institute for Modeling the Earth System, Princeton University (Award NA18OAR4320123 from the National Oceanic and Atmospheric Administration, U.S. Department of Commerce), and by the French Agence National de Recherche ``Investissements d'avenir'' \emph{Make Our Planet Great Again} program (ANR-17-MPGA-0010). We thank Naser Mahfouz, Zhihong Tan, and two anonymous reviewers for substantive and detailed reviews that have immeasurably improved the manuscript.

\bibliography{gcmobsolete}

\end{document}